\documentclass{iaus}
\usepackage{graphicx}

\title[IAUS 271.~~Dynamo action in B stars]{Convection and dynamo action in B stars}

\author[Augustson et al.]{Kyle C. Augustson$^1$, Allan S. Brun$^2$, \and Juri Toomre$^1$}

\affiliation{$^1$JILA \& APS, University of Colorado, 440 UCB, Boulder, CO, 80309-0440, USA \\ [\affilskip]
$^2$UMR AIM CEA-CNRS-Univ. P7 CEA Saclay,  91191 Gif-sur-Yvette Cedex, France}

\pubyear{2010}
\volume{271}
\pagerange{100--101}
\jname{Astrophysical Dynamics: From Galaxies to Stars}
\editors{N.H. Brummell \& A.S. Brun}

\def\apj{\textit{ApJ}}
\def\araa{\textit{ARAA}}

\def\pre{\textit{Phys. Rev. E}}
\def\etal{et al.}
\def\sun{\odot}
\def\msun{$\mathrm{M_{\sun}}$}
\def\lsun{$\mathrm{L_{\sun}}$}
\def\osun{$\mathrm{\Omega_{\sun}}$}
\def\rstar{$\mathrm{R_{*}}$}
\begin{document}

\maketitle

\begin{abstract}
Main-sequence massive stars possess convective cores that likely harbor strong dynamo 
action. To assess the role of core convection in building magnetic fields within these 
stars, we employ the 3-D anelastic spherical harmonic (ASH) code to model turbulent 
dynamics within a 10~\msun\ main-sequence (MS) B-type star rotating at 4~\osun.
We find that strong (900~kG) magnetic fields arise within the turbulence of the core 
and penetrate into the stably stratified radiative zone. These fields exhibit complex, 
time-dependent behavior including reversals in magnetic polarity and shifts between which 
hemisphere dominates the total magnetic energy.
\end{abstract}

Surface magnetic fields have been found on many MS massive stars (e.g. \cite[Donati \& Landstreet 2009]{donati09}).
To some degree both the fossil fields and dynamo generated fields in these stars must coexist, 
although how they interact to produce and maintain these surface fields is unclear. Recent work 
has shed some light on the interaction between a super-equipartition core dynamo and fossil magnetic 
fields in A-type stars (\cite[Brun \etal\ 2005]{brun05}, \cite[Featherstone \etal\ 2009]{featherstone09}). 
We extend this work to a much more luminous 10~\msun\ star with a rotation period of seven
days (4~\osun) which is typical for active MS B stars.

Using the 3-D ASH code, we study convection and dynamo action realized in the core and part of the 
surrounding radiative envelope of this 7200~\lsun\ B star. ASH is a mature modeling 
tool which solves the anelastic MHD equations of motion in a rotating spherical shell using a 
pseudo-spectral method (e.g. \cite[Brun \etal\ 2004]{brun04}). The mean structure in this ZAMS star 
is obtained from a stellar evolution code. We capture the full spherical geometry with a radial domain 
that occupies 0.6~\rstar\ (covering 7 pressure scale heights), with the inner 0.2~\rstar\ being 
convectively unstable. The innermost 0.02~\rstar\ is excluded to avoid the coordinate singularity at 
the origin in the ASH code. The upper and lower radial boundary conditions are stress-free and impenetrable 
for the velocity field and perfect conductor (lower) and potential field (upper) for the magnetic field.

\begin{figure}[t]
\begin{center}
\includegraphics[width=\textwidth]{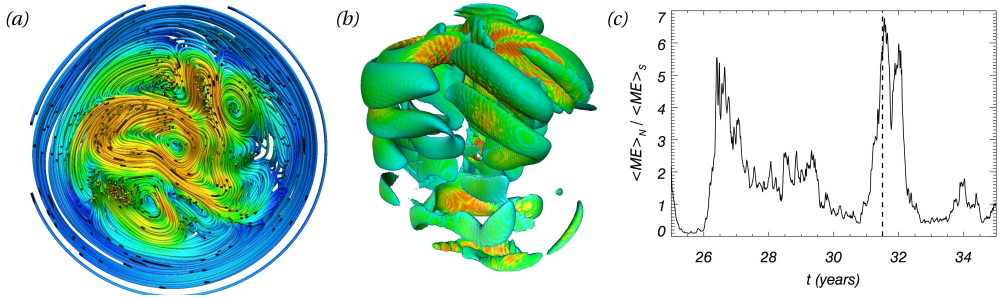}
\caption{(a) Velocity streamlines within the core, cut along the equator. Fast core-crossing 
  flows and several columnar flows are visible. Orange tones indicate fast flow speeds 
  ($|v| > 300 \, \mathrm{m\,s^{-1}}$, peak 1000~$\mathrm{m\,s^{-1}}$), slower speeds in blue tones. (b) An 
  isocontour rendering of magnetic energy showing the dominance of the northern hemisphere and equatorward 
  tilted magnetic structures (rotation axis vertical). (c) North to south ratio of hemispherical averages 
  of ME shown for a decade of time evolution; (a) and (b) are rendered at 31.5 years (dashed line).}
\end{center}
\vspace{-0.07truein}
\end{figure}

The intricate and time-varying flows established in this simulation are largely aligned with the rotation 
axis. These columnar convection cells break the spherical symmetry due to equator-crossing meridional 
circulations and a north-south asymmetric differential rotation. A central columnar flow (occupying the 
inner 0.1~\rstar\ at the equator) stretches north-south across the entire core, rotates retrograde to 
the reference frame, and gently flares out to about $25^{\circ}$ in latitude at the core boundary (Fig. 1a). 
Along the rotation axis within this column are strong vortical flows. Outside the central column, there 
are typically five columnar convection cells that rotate prograde to the reference frame. These cells 
transport angular momentum between the central column and the overshooting region, where there is a weak 
prograde equatorial flow. These flows maintain a mean rotation rate that increases monotonically from the 
center of the star to become nearly uniform within the radiative envelope, with an overall radial 
differential rotation of 25\%.

A strong dynamo operates within the core, generating magnetic fields with peak strengths reaching 900~kG
(200~kG rms). These fields form equatorward tilted strands that encircle the core (Fig. 1b). 
The fluctuating component of the magnetic field comprises 76\% of the total magnetic energy (ME) in the 
core, while 21\% remains in the mean toroidal field and 3\% in the mean poloidal field. On average 
the total ME is 55\% of the convective kinetic energy, but there are intervals where it approaches 86\% 
indicating that the ME is nearing equipartition. 

The time evolution of magnetic field is complex and multi-periodic. When averaged over several decades, 
the ME of the northern hemisphere is 1.7 times greater than that of the southern hemisphere. There are 
intervals, however, when the southern hemisphere comes to dominate the magnetic energy (Fig. 1c), but only 
by a factor of at most 2.5. The northern hemisphere, on the other hand, dominates the ME for periods of up 
to two years by a factor as great as 6.7. These magnetic field configurations have quadrupolar and dipolar 
components that are nearly equal and opposite, which have been shown to exist when there is weak equatorial 
symmetry breaking (\cite[Gallet \& P\'{e}tr\'{e}lis 2009]{gallet09}).

The greatest extent of convective overshooting into the stable radiative envelope occurs at mid-latitudes. 
The sustained overshooting pushes magnetic field and lower entropy fluid into the stable layer, making the 
core prolate and stochastically exciting gravity waves. The strongest magnetic fields (900~kG) and fastest 
flows (1~$\mathrm{km\,s^{-1}}$) typically occur along the edge of the central column and are maximum where 
this column transects the core boundary. As this field is advected into the overshooting region it is combed 
into a large-scale toroidal field ($\sim$30~kG) by the flows in the stable region. Therefore, in this region, 
the velocity and magnetic fields are nearly aligned creating a force-free state.

To better understand the hemispherical dynamo state achieved within this B-star model, simulations at
varying rotation rates and lower diffusivities must be run. Minimal diffusion is especially important
if we are to capture the buoyant magnetic structures that likely arise from the strongest fields in 
these models. 

This work was supported by NASA grants NNX08AI57G and NNX10AM74H and by NSF Teragrid supercomputing resource 
grant TG-MCA93S005.


\vspace{-0.1truein}
\begin{thebibliography}{}
\small

\bibitem[Brun \etal\ (2004)]{brun04}
{{Brun}, A.S., {Miesch}, M.S., \& {Toomre}, J.} 2004,
{\apj}, 614, 1073

\bibitem[Brun \etal\ (2005)]{brun05}
{{Brun}, A.S., {Browning}, M.K., \& {Toomre}, J.} 2005,
{\apj}, 629, 461

\bibitem[Donati \& Landstreet (2009)]{donati09}
{{Donati}, J.-F. \& {Landstreet}, J.D.} 2009, {\araa}, 47, 333

\bibitem[Featherstone \etal\ (2009)]{featherstone09}
{{Featherstone}, N.A., {Browning}, M.K., {Brun}, A.S., \& {Toomre}, J.} 2009, {\apj}, 705, 1000

\bibitem[Gallet \& P\'{e}tr\'{e}lis (2009)]{gallet09}
{{Gallet}, B., \& {P\'{e}tr\'{e}lis}, F.} 2009, {\pre}, 80, 3

\end{thebibliography}
\end{document}